\documentclass{emulateapj}
\usepackage{graphics}
\usepackage{color}
\usepackage{amsmath}
\usepackage{empheq} 
\newcommand{\gtsimeq}{\ga}

\newcommand{\HI}{H{\sc i}}
\newcommand{\HII}{H{\sc ii}}

\received{}
\revised{}
\accepted{}
\shortauthors{McQuinn et al.}
\shorttitle{Accurate Distances to Important Spiral Galaxies}

\begin{document}
\title{Accurate Distances to Important Spiral Galaxies: M63, M74, NGC~1291, NGC~4559, NGC~4625, NGC~5398}\thanks{Based on observations made with the NASA/ESA Hubble Space Telescope, obtained from the Data Archive at the Space Telescope Science Institute, which is operated by the Association of Universities for Research in Astronomy, Inc., under NASA contract NAS 5-26555.}
\author{Kristen~B.~W. McQuinn\altaffilmark{1,2}, 
Evan D. Skillman\altaffilmark{2},
Andrew E.~Dolphin\altaffilmark{3},
Danielle Berg\altaffilmark{4}, 
Robert Kennicutt\altaffilmark{5}
}

\altaffiltext{1}{University of Texas at Austin, McDonald Observatory, 2515 Speedway, Stop C1400 Austin, Texas 78712, USA \ {\it kmcquinn@astro.as.utexas.edu}}
\altaffiltext{2}{Minnesota Institute for Astrophysics, School of Physics and Astronomy, 116 Church Street, S.E., University of Minnesota, Minneapolis, MN 55455, USA} 
\altaffiltext{3}{Raytheon Company, 1151 E. Hermans Road, Tucson, AZ 85756, USA}
\altaffiltext{4}{Center for Gravitation, Cosmology and Astrophysics, Department of Physics, University of Wisconsin Milwaukee, 1900 East Kenwood Boulevard, Milwaukee, WI 53211, USA}
\altaffiltext{5}{Institute for Astronomy, University of Cambridge, Madingley Road, Cambridge CB3 0HA, England }

\begin{abstract}
Accurate distances are fundamental to interpreting many measured properties of galaxies. Surprisingly, many of the best-studied spiral galaxies in the Local Volume have distance uncertainties that are much larger than can be achieved with modern observations. Using Hubble Space Telescope optical imaging, we use the tip of the red giant branch method to measure the distances to six galaxies that are included in the Spitzer Infrared Nearby Galaxies Survey (SINGS) program and its offspring surveys. The sample includes M63, M74, NGC~1291, NGC~4559, NGC~4625, and NGC~5398. We compare our results with distances reported to these galaxies based on a variety of methods. Depending on the technique, there can be a wide range in published distances, particularly from the Tully-Fisher relation. In addition, differences between the Planetary Nebula Luminosity Function and Surface Brightness Fluctuation techniques can vary between galaxies suggesting inaccuracies that cannot be explained by systematics in the calibrations. Our distances improve upon previous results as we use a well-calibrated, stable distance indicator, precision photometry in an optimally selected field of view, and a Bayesian Maximum Likelihood technique that reduces measurement uncertainties. 
\end{abstract} 


\section{Introduction}\label{sec:intro}
Great investments of observing time have been dedicated to the study of nearby spiral galaxies with diverse goals ranging from understanding the star formation process to characterizing their dark matter distributions. Accurate distances are fundamental to interpreting observations of these galaxies, yet many of the best studied nearby galaxies have distances based on methods with relatively large uncertainties. 

We have undertaken a survey to provide precise distance measurements using the tip of the red giant branch (TRGB) method for a sample of nearby galaxies that lack secure distances. The full sample includes the famous spiral galaxies the Whirlpool (M51; NGC~5194), the Sunflower (M63; NGC~5055), and M74 (NGC~628; the archetype grand-design spiral), the Sombrero (M104; NGC~4555) as well as the four spiral galaxies NGC~5398, NGC~1291, NGC~4559, NGC~4625.  Our sample is part of many detailed surveys of nearby galaxies including the Spitzer Infrared Nearby Galaxies Survey \citep[SINGS;][]{Kennicutt2003}, the GALEX Space Telescope Nearby Galaxy Survey  \citep[NGS][]{GildePaz2007}, The \HI\ Nearby Galaxy Survey (THINGS) with the VLA \citep{Walter2008}, the IRAM 30-m telescope HERA CO Line Extragalactic Survey (HERACLES) project\citep{Leroy2009}, the Key Insights on Nearby Galaxies: a Far-Infrared Survey with Herschel (KINGFISH) program \citep{Kennicutt2011}, the PPAK optical Integral field spectroscopy Nearby Galaxies Survey \citep[PINGS][]{Rosales2010}, and the optical slit-let spectroscopy CHemical Abundances of Spirals (CHAOS) program \citep{Berg2015}. 

\begin{figure*}
\begin{center}
\includegraphics[width=0.97\textwidth]{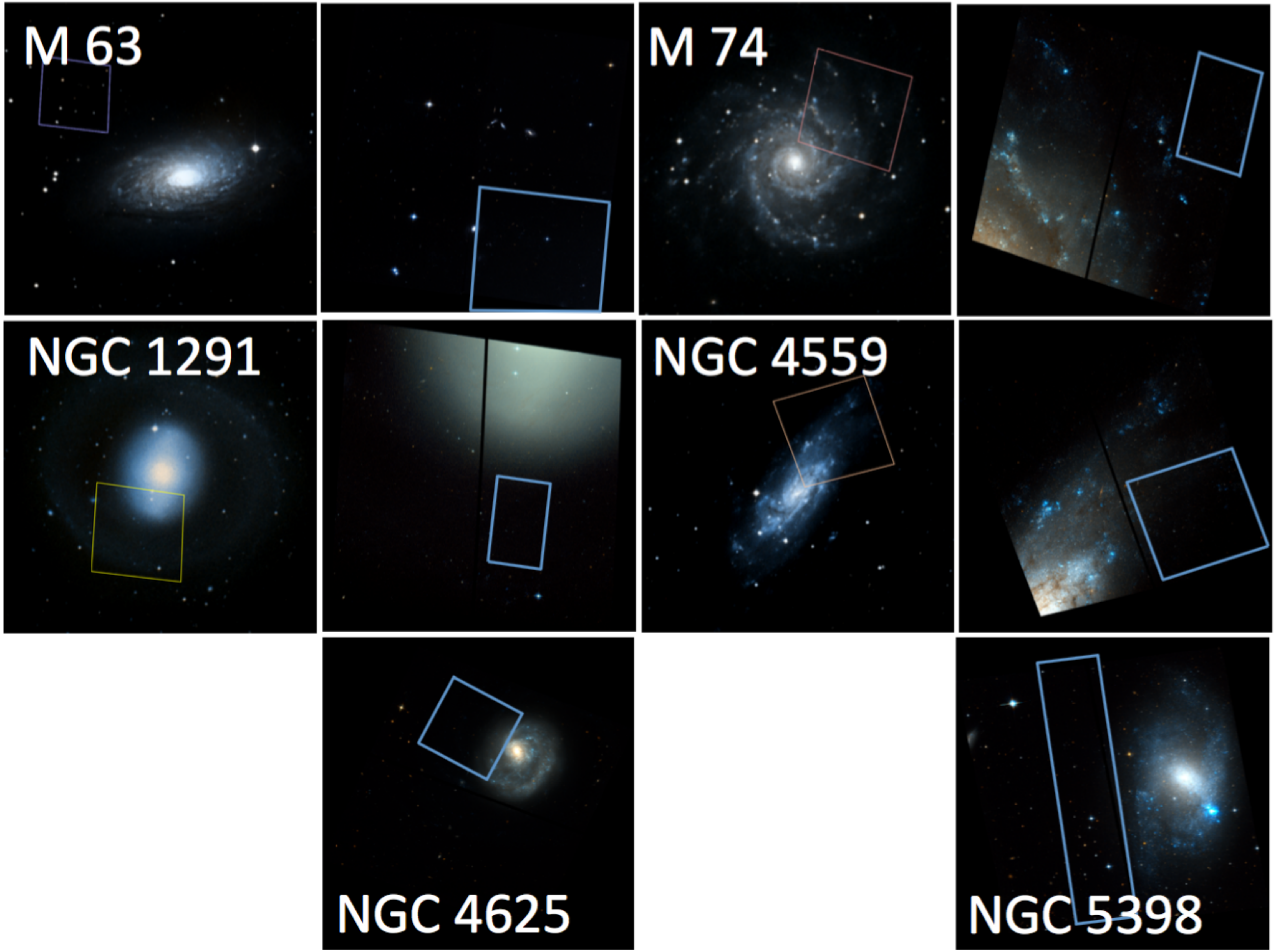}
\end{center}
\caption{{\it Left panels for each galaxy:} DSS image of individual galaxies overlaid with the $HST$ field of view of our observations. {\it Right panels for each galaxy:} $HST$ ACS imaging of our fields created by combining F606W (blue), F814W (red), and an average of the two filters (green). The images are oriented with North up and East left. DSS images are not given for NGC~4625 or NGC~5298 as the ACS images cover the main stellar disks of the galaxies. Regions selected for the TRGB analysis are overlaid in blue on the ACS images.}
\label{fig:images}
\end{figure*}

%%tab:properties
\begin{table*}
%McQuinn et al. 2017 Distances to Important Galaxies Table 1
\begin{center}
\caption{Galaxy Sample and Observations}
\label{tab:properties}
\end{center}
\begin{center}
\vspace{-15pt}
\begin{tabular}{l | cccccc}
\hline 
\hline 
Galaxy		& RA 		& Dec		& $A_{F606W}$& $A_{F814W}$& F606W	& F814W \\
			& (J2000)		& (J2000)		& (mag)		& (mag)		& (sec)	& (sec) 	\\
\hline
M63			& $13:15:49.3$ & $+42:01:45$	& 0.043		& 0.027		& 2168	& 2168 \\
M74			& $01:36:41.8$ & $+15:47:01$	& 0.173		& 0.107		& 2553	& 2553 \\
NGC~1291	& $01:36:41.8$	& $-41:06:29$	& 0.032		& 0.020		& 2639	& 2639 \\
NGC~4559	& $12:35:57.7$	& $+27:57:36$	& 0.043		& 0.027		& 2632	& 2632 \\
NGC~4625	& $12:41:52.7$	& $+41:16:26$	& 0.045		& 0.028		& 2639	& 2639 \\
NGC~5398	& $14:01:21.6$	& $-33:03:49$	& 0.163		& 0.100		& 2579	& 2579 \\
\hline
\end{tabular}
\end{center}
\tablecomments{Coordinates, extinction, and integration times for the sample. Observation times are from program GO$-$13804 (PI McQuinn) for all galaxies except M63 (GO$-$10904; PI Thilker). Galactic extinction estimates are from \citet{Schlafly2011}.}

\end{table*}

The first two papers in this series focused on measuring the TRGB distance to M51 \citep[D$=8.58\pm0.10$ (statistical) Mpc;][hereafter Paper~I]{McQuinn2016a} and M104 \citep[D$=9.55\pm0.13\pm0.31$ Mpc;][hereafter Paper~II]{McQuinn2016b}. In Paper~I, we described the observation strategy, data analysis, and methodology in detail; in the Appendices of Papers~I and II we provided descriptions of the various previous methods used to measure the distances to the two galaxies.. 

Here, we report the TRGB distances for the remainder of the sample; we include the distances to M51 and M104 in our final discussion for completeness. In Section~\ref{sec:trgb}, we present a summary of the TRGB distance method. In Section~\ref{sec:obs} we provide a summary of the observations and data processing. In Section~\ref{sec:distances} we discuss the measurement of the TRGB distances. In Section~\ref{sec:compare}, we present a comparison of previous distances from the literature for each galaxy. In the Appendix, we provide descriptions of two additional distance methods not discussed in Papers~I and II. In Section~\ref{sec:conclusions}, we summarize our conclusions.  

\section{Precise and Efficient TRGB Distances}\label{sec:trgb}
The TRGB method is one of the most precise and efficient approaches to measuring distances in the nearby universe. The TRGB method is a standard-candle distance indicator based on the predictable maximum luminosity of red giant branch (RGB) stars just prior to the helium flash \citep{Mould1986, Freedman1988, DaCosta1990, Lee1993}. Low-mass stars evolve up the RGB burning hydrogen in a shell with an electron-degenerate helium core and a convective outer envelope. The luminosity of an RGB star increases until helium burning in the core ignites in a He-flash and the star evolves off the RGB. The transition to helium burning occurs at a predictable luminosity  in the $I-$band corresponding to the TRGB, independent of stellar mass, with a modest dependence on stellar metallicity \citep[e.g.,][]{Lee1993, Salaris1997}. 

The TRGB distance method is based on identifying the discontinuity in the $I-$band luminosity function (LF) of stars in the upper RGB. Thus, imaging of the resolved stellar populations is needed in the $I-$band reaching $\gtsimeq1$ mag below the expected TRGB and in a bluer optical wavelength to allow the RGB stars to be selected from a composite stellar population based on color criteria. The TRGB luminosity is well-calibrated at optical wavelengths \citep[e.g.,][]{Bellazzini2001} and to Hubble Space Telescope (HST) filters \citep{Rizzi2007a}, which include corrections for the metallicity dependence. TRGB distances can be measured for galaxies in and around the Local Group from ground-based imaging, and out to $\sim10$ Mpc using single orbit observations with the high resolution and sensitivity of $HST$ optical imaging instruments. 

\section{Observations and Photometry}\label{sec:obs}
Table~\ref{tab:properties} lists the sample with their coordinates, foreground extinction, and observation details. Observations for five of the galaxies are from the HST-GO-13804 program (PI: McQuinn); observations for M63 are archival from HST-GO-10904 (PI: Thilker). All observations were obtained with the $HST$ using the Advanced Camera for Surveys (ACS) Wide Field Channel (WFC) \citep{Ford1998}. Images were taken with ACS during two $HST$ orbits in the F606W and F814W filters with integration times of $\sim2500$ s per filter. For our observations, we used a standard 2-point hot-pixel dither pattern to reduce the impact of bad pixels and cosmic rays. Similarly, the archival images were taken using a 2-point dither pattern and were cosmic-ray split (CR-SPLIT). The data were corrected for the effects of charge transfer efficiency (CTE) non-linearities caused by space radiation damage on the ACS instrument \citep[e.g.,][]{Anderson2010, Massey2010} and processed by the standard ACS data pipeline. 

Figure~\ref{fig:images} presents images of the sample. Four galaxies have an angular extent significantly greater than the ACS field of view. For these galaxies, we present DSS images with the ACS field of view overlaid (left panels) alongside the ACS images (right panels). For the remaining two systems with smaller angular extent, NGC~4625 and NGC~5398, we present only the ACS images. The 3-color ACS images were made by combining F606W (blue), the average of F606W and F814W (green) and F814W (red) drizzled images created from CTE corrected images (flc.fits files) using DrizzlePac 2.0 \textsc{Astrodrizzle} software. 

In Figure~\ref{fig:images}, the images show a range of structure, surface brightness, and stellar crowding. As described in Paper~I in detail, the chosen fields of view for the observations targeted the outer disks of the galaxies. These outer fields have a number of advantages for measuring the TRGB luminosity over inner regions. First, they are less crowded, which improves photometric accuracy. Second, the RGB stars typically have lower metallicity, which ensures higher photometric completeness at the expected F606W-F814W colors of $1-2$ mag of the RGB stars given our integration times. Third, the general age of the stellar populations in the outer regions is older, which reduces contamination in the CMD from intermediate-age thermally-pulsating asymptotic giant branch (TP-AGB) stars. The spiral galaxies are massive enough that even in these outer regions there are large number of RGB stars in each ACS field of view.The exception is M63 where the archival images are at a great enough galactocentric distance that the ACS fields seem under-populated relative to the remainder of the sample. However, this outer field still contains tens of thousands of stars, more than a sufficient number for a TRGB distance determination. 

In selecting the fields of view, we also chose orientations for the ACS instrument such that the Wide Field Camera 3 (WFC3) instrument imaged star-forming regions in the ultraviolet simultaneously in a parallel observing mode. The WFC3 imaging will be discussed in a future paper.

\begin{figure*}
\includegraphics[width=\textwidth]{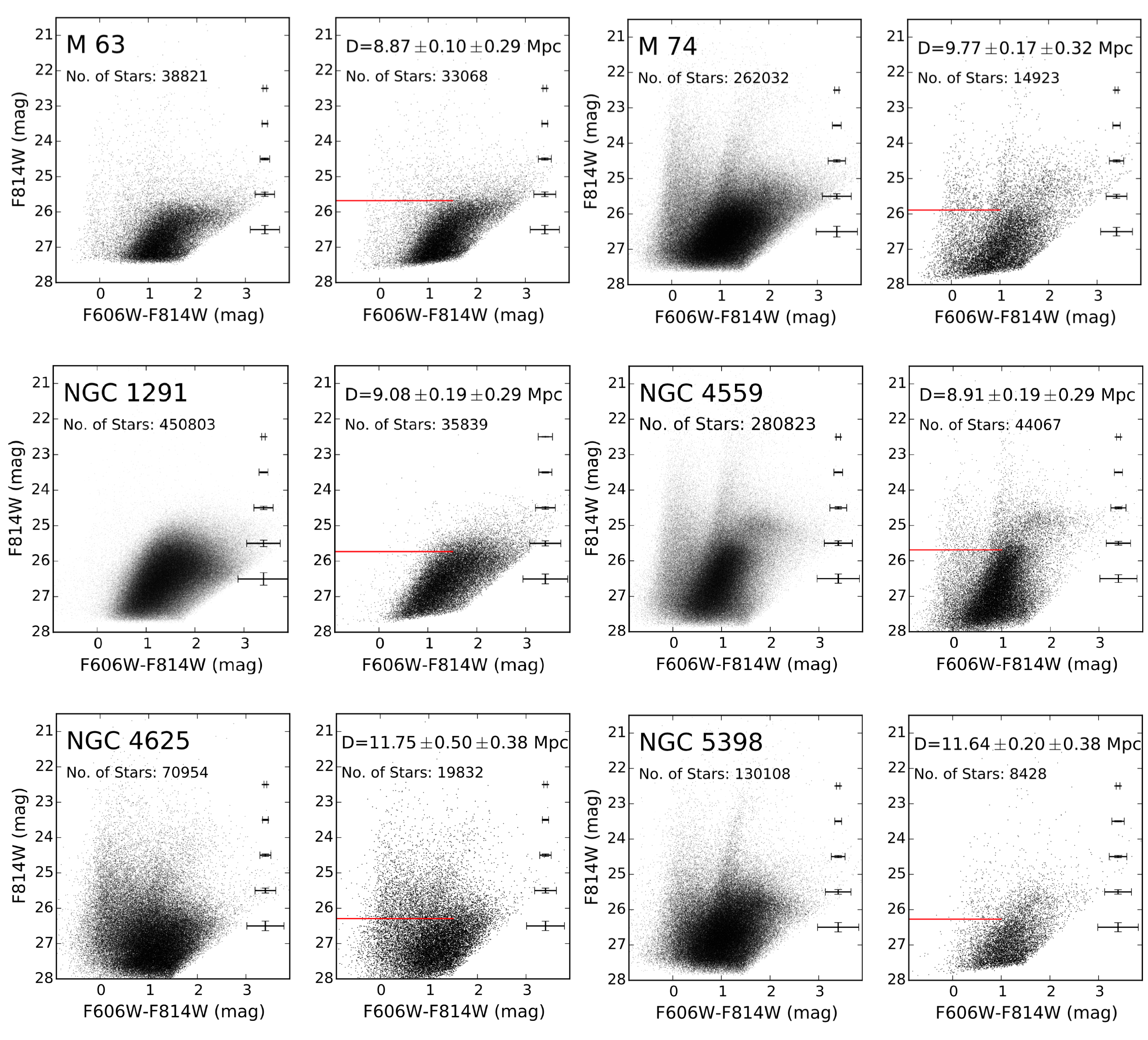}
\caption{{\em Left panels for each galaxy:} CMDs of the full $HST$ fields of view. {\em Right panels for each galaxy:} CMDs of the regions selected for TRGB analysis. All CMDs have been corrected for foreground extinction. The uncertainties are from the photometry and artificial star tests. Uncertainties in color are larger due to the inclusion of F606W photometry with low SNRs. The CMDs in the right panels have photometry that were transformed using the color-based calibration correction for metallicity from \citet{Rizzi2007a}. By applying this correction before fitting for the TRGB (instead of applying the correction in the final calibration), the curvature in the RGB is reduced allowing for the TRGB to be measured with a higher degree of certainty. The TRGBs measured for each galaxy are marked with a red horizontal line.}
\label{fig:cmd}
\end{figure*}

PSF photometry on the ACS CTE-corrected images was done using DOLPHOT, a modified version of HSTphot optimized for the ACS instrument \citep[][]{Dolphin2000}. Artificial star tests were also performed on the images using the same photometry software. The resulting photometric and artificial star lists were filtered for high fidelity point sources based on a number of criteria including signal-to-noise ratio (SNR) thresholds, sharpness cuts, and crowding limits. For the F814W photometry, we required a SNR minimum of 5$\sigma$, whereas for the F606W photometry we set a lower threshold of 2$\sigma$. This lower threshold increases the number of the point sources at fainter F606W magnitudes, which more fully populates the bottom of the CMD at the typical colors of RGB stars. The inclusion of F606W point sources with a lower SNR prevents introducing an artificial discontinuity in the F814W LF.

Photometry was also filtered to reject point sources with high measured sharpness values (i.e., sources whose flux was sharply peaked indicating the source was likely a cosmic ray) or a low sharpness value (i.e., sources with a broad PSF indicating the source is likely a background galaxy). Finally, point sources in crowded regions were also rejected. Specifically, we rejected point sources with (V$_{sharp}+$I$_{sharp}$)$^2>0.075$ and (V$_{crowd}+$I$_{crowd}$)$>0.8$. The photometric catalogs were corrected for foreground Galactic extinction using the dust maps from \citet{Schlegel1998} and the calibration from \citet{Schlafly2011}.

Figure~\ref{fig:cmd} presents the extinction corrected CMDs for the sample, plotted to the $\sim$50\% completeness level. The left panel for each galaxy shows the CMD from the full ACS field of view; each CMD has a clearly identifiable RGB reaching $\sim$2 mag below the TRGB. Representative uncertainties per magnitude based on combining photometric uncertainties and the results from the artificial star tests are shown. The uncertainties in color at fainter magnitudes are larger as we included F606W sources with lower SNRs. 

From Figure~\ref{fig:cmd}, it is readily apparent that the observed fields of view have large numbers of stars and contain composite stellar populations. As noted above, the ACS fields of view encompass a wide range in structure, stellar density, and surface brightness. Because identifying the TRGB is based on measuring the discontinuity in the F814W LF, the inclusion of non-RGB stars in the photometry with colors and luminosities consistent with RGB stars can broaden the break in the LF and increase the statistical uncertainties. Therefore, we selected regions in the ACS images for the TRGB measurements at galactocentric radii that avoided significant star-forming complexes. These regions are typically at larger galactocentric radii in the ACS images and have higher levels of photometric completeness at the approximate magnitude of the TRGB. In the case of M74, we also compared the radial distribution of stars above and below the approximate TRGB identified by eye in the CMD, representing an approximate separation of AGB and RGB stars. This coarse comparison helped guide our region selection to an area that contained fewer sources brighter than the region of the TRGB. In the case of M63, we chose the region that was closest in galactocentric radii to reduce background contamination in the CMD from the relatively underpopulated outer region. In Figure~\ref{fig:images}, we also overlay in blue the regions selected for the TRGB analysis on the ACS images. 

The right panels in Figure~\ref{fig:cmd} present CMDs from the regions selected for the TRGB measurements. The photometry includes $\sim10$k$-40$k stars per region with less contamination from composite stellar populations. Photometric completeness in the TRGB magnitude range measured by artificial star tests increased by $\sim10-15$\% in the selected regions compared to the full field of view.

In the F814W filter, the luminosity of the TRGB has a well-known, modest dependency on metallicity. TRGB calibrations can account for this dependency by using a color-based correction. We adopt the calibration from \citet{Rizzi2007a} derived for the $HST$ filters, reproduced here for convenience:

\begin{equation}
M_{F814W}^{ACS} = -4.06+ 0.20 \cdot [(F606W-F814W) - 1.23] \label{eq:trgb}
\end{equation}

\begin{figure*}
\includegraphics[width=0.98\textwidth]{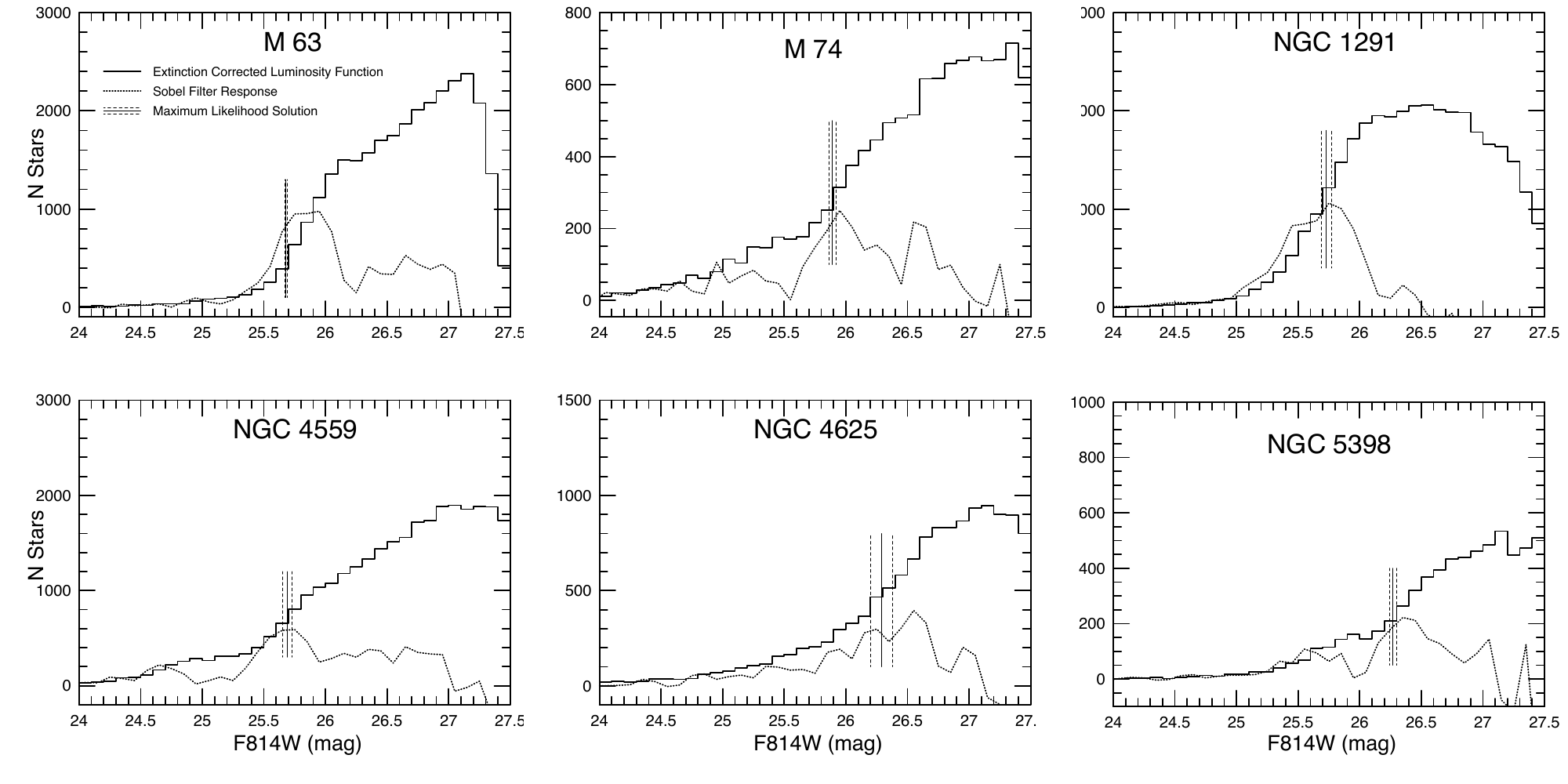}
\caption{Luminosity functions for the F814W photometry corrected for Galactic extinction and transformed with the color-based metallicity correction. The Sobel filter responses are overplotted as dotted lines. The TRGB luminosity and $1\sigma$ statistical uncertainties measured by the ML technique are marked with vertical solid and dashed lines respectively.}
\label{fig:sobel}
\end{figure*}

From Equation~\ref{eq:trgb}, the metallicity correction uses the average  TRGB $F606W - F814W$ color to account for the difference between the metallicity of the galaxies used in the calibration and the target galaxy. While one can measure the TRGB magnitude and then apply the metallicity correction and zeropoint, we chose to apply the metallicity correction to the photometry prior to identifying the TRGB. Thus, the overall curvature in the RGB is reduced, increasing the sharpness in the discontinuity from the TRGB in the LF. 

The CMDs in the right panels of Figure~\ref{fig:cmd} include photometry that has been transformed by the color-correction in Equation~\ref{eq:trgb}. Comparing the left and right panels, the RGBs have less curvature and the TRGB features in the CMDs appear flatter. We use the extinction corrected and transformed photometry for the TRGB measurement. Once identified, the zeropoint from Equation~\ref{eq:trgb} is applied to the TRGB magnitudes for the final, calibrated distance moduli.

\section{Measuring the TRGB Discontinuity in the F814W Luminosity Function}\label{sec:distances}
We use two techniques to identify and measure the discontinuity in the F814W LF. The first is a Sobel filter edge detection technique \citep{Lee1993, Sakai1996, Sakai1997}, which is based on applying a Sobel kernel in the form [$-2$, 0, $+2$] to a binned LF. The second is a Bayesian Maximum Likelihood (ML) technique which fits the observed LF function to a parametric representation of the RGB stars LF. The resulting probability estimation takes into account the photometric error distribution and completeness function from artificial star tests \citep[see][for a full discussion]{Makarov2006}. For the ML technique, we used the following form for the theoretical LF:

\begin{subequations}
\begin{empheq}[left={P = }\empheqlbrace]{alignat=2}
        & 10^{(A*(m - m_{TRGB}) + B)}, & \quad \text{if m - m$_{TRGB} \geq 0$}\\
        & 10^{(C*(m - m_{TRGB}))}, & \quad \text{if m - m$_{TRGB} < 0$}
\end{empheq}
\label{eq:ml_form}
\end{subequations}

\noindent where A is the slope of the RGB with a normal prior of 0.30 and $\sigma=0.07$, B is the RGB discontinuity, and C is the slope of the AGB with a normal prior of 0.30 and $\sigma=0.2$. All variables are treated as free parameters. This is the same theoretical LF form used in \citet{Makarov2006}. The range in solutions returning log(P) within 0.5 of the maximum gives the uncertainty (as is the case with a normal distribution). 

%tab:distances
\begin{table*}
%McQuinn et al. 2017 Distances to Important Galaxies Table 2
\begin{center}
\caption{Summary of Distances}
\label{tab:distances}
\end{center}
\begin{center}
\vspace{-15pt}
\begin{tabular}{l | cccc}
\hline 
\hline 
Galaxy		& RGB Colors		& TRGB$_0$ 		& Distance Modulus			& Distance \\
			& (F606W-F814W)	& (mag)			& (mag)					& (Mpc)	\\
\hline
M51			& $0.5-2.0$		& $25.61\pm0.01$	& $29.67\pm0.02\pm0.07$	& $8.58\pm0.10\pm0.28$ \\
M63			& $0.6-2.5$		& $25.68\pm0.01$	& $29.74\pm0.02\pm0.07$	& $8.87\pm0.10\pm0.29$ \\
M74			& $0.5-2.4$		& $25.89\pm0.03$	& $29.95\pm0.04\pm0.07$	& $9.77\pm0.17\pm0.32$\\
M104		& $0.5-2.6$		& $25.84\pm0.02$	& $29.90\pm0.03\pm0.07$	& $9.55\pm0.13\pm0.31$ \\
NGC~1291	& $0.5-2.0$		& $25.73\pm0.04$	& $29.79\pm0.05\pm0.07$	& $9.08\pm0.19\pm0.29$ \\
NGC~4559	& $0.5-2.0$		& $25.69\pm0.04$	& $29.75\pm0.05\pm0.07$	& $8.91\pm0.19\pm0.29$ \\
NGC~4625	& $0.5-3.0$ 		& $26.29\pm0.09$	& $30.35\pm0.09\pm0.07$	& $11.75\pm0.50\pm0.38$\\
NGC~5398	& $0.7-2.3$		& $26.27\pm0.03$	& $30.33\pm0.04\pm0.07$	& $11.64\pm0.20\pm0.38$ \\
\hline
\end{tabular}
\end{center}
\tablecomments{RGB colors refer to limits applied to reduce inclusion of non-RGB stars when fitting for the TRGB discontinuity in the F814W LF. The TRGB magnitude is measured from the extinction corrected photometry after applying the color-based metallicity calibration correction. Statistical and systematic uncertainties are separately listed for the distance moduli and distances. Measurements for M51 and M104 are from \citet{McQuinn2016a} and \citet{McQuinn2016b} respectively. They are repeated here for completeness.}

\end{table*}

Generally, the ML technique is preferred over the Sobel filter approach because the ML technique both avoids binning the LF and considers the photometric uncertainties and completeness. In addition, the ML technique fits the F814W LF while taking into account the presence of an AGB population. Thus, ML statistical uncertainties are typically lower than those measured from the Sobel filter edge detection determined mainly from the bin size. However, the Sobel filter edge detection approach is useful in providing a consistency check on the ML results. 

We use the extinction corrected, transformed photometry to select point sources with luminosities and colors consistent with RGB stars for each galaxy. By reducing the number of non-RGB stars used in the analysis, we reduce the Poisson noise in the TRGB fits. The average color and width of the RGB is unique for each system and therefore must be determined on a galaxy by galaxy basis. We chose broad color limits (listed in Table~\ref{tab:distances}) to ensure we are fitting the full TRGB population with the maximum number of stars. 

Figure~\ref{fig:sobel} presents the F814W LF of the stars used in the distance measurements with the Sobel filter response overlaid. In most cases, the LF shows a more subtle change rather than a large, sharp discontinuity at the magnitude of the TRGB, and the corresponding Sobel response has a broad profile rather than being sharply peaked. This is consistent with the distribution of the stars seen in the CMDs in Figure~\ref{fig:cmd}, where there is a population of point sources brighter than the TRGB identifiable by eye. In Figure~\ref{fig:sobel}, we overlay the TRGB magnitude identified by the ML technique, with statistical uncertainties from the fit. These fits are consistent with the broad peaks of the Sobel filter response. We adopt the ML TRGB magnitudes and uncertainties for our distance calculations.

Table~\ref{tab:distances} lists the best-fitting value for the TRGB from the ML techniques for the sample; uncertainties are the statistical uncertainties from the ML probability distribution function. Each TRGB value is also marked in the CMDs in Figure~\ref{fig:cmd}. The TRGB luminosities are converted to distance moduli using the zero-point calibration of $-4.06$ from \citet{Rizzi2007a}. We combine the statistical uncertainties of our measurements with the statistical uncertainties from the calibration for TRGB zero-point  ($\sigma=$ 0.02), and color-dependent metallicity correction  ($\sigma=$ 0.01). We adopt the systematic uncertainties from \citet{Rizzi2007a} of 0.07 mag, but note that higher systematic uncertainties have been reported for TRGB calibrations that are based on $\omega$ Centauri distance measurements \citep[e.g., $\sigma=$0.12;][]{Bellazzini2001}. The final distances with statistical and systematic uncertainties are listed in Table~\ref{tab:distances}. We include the values for M51 \citep{McQuinn2016a} and M104 \citep{McQuinn2016b} for completeness. For the distance measurement of M51, we did not report a systematic uncertainty in Paper~I. Here, we adopt the same systematic uncertainties from \citet{Rizzi2007a}.

\section{Comparison with Previous Distances for Each Galaxy}\label{sec:compare}
\begin{figure}
\includegraphics[width=\linewidth]{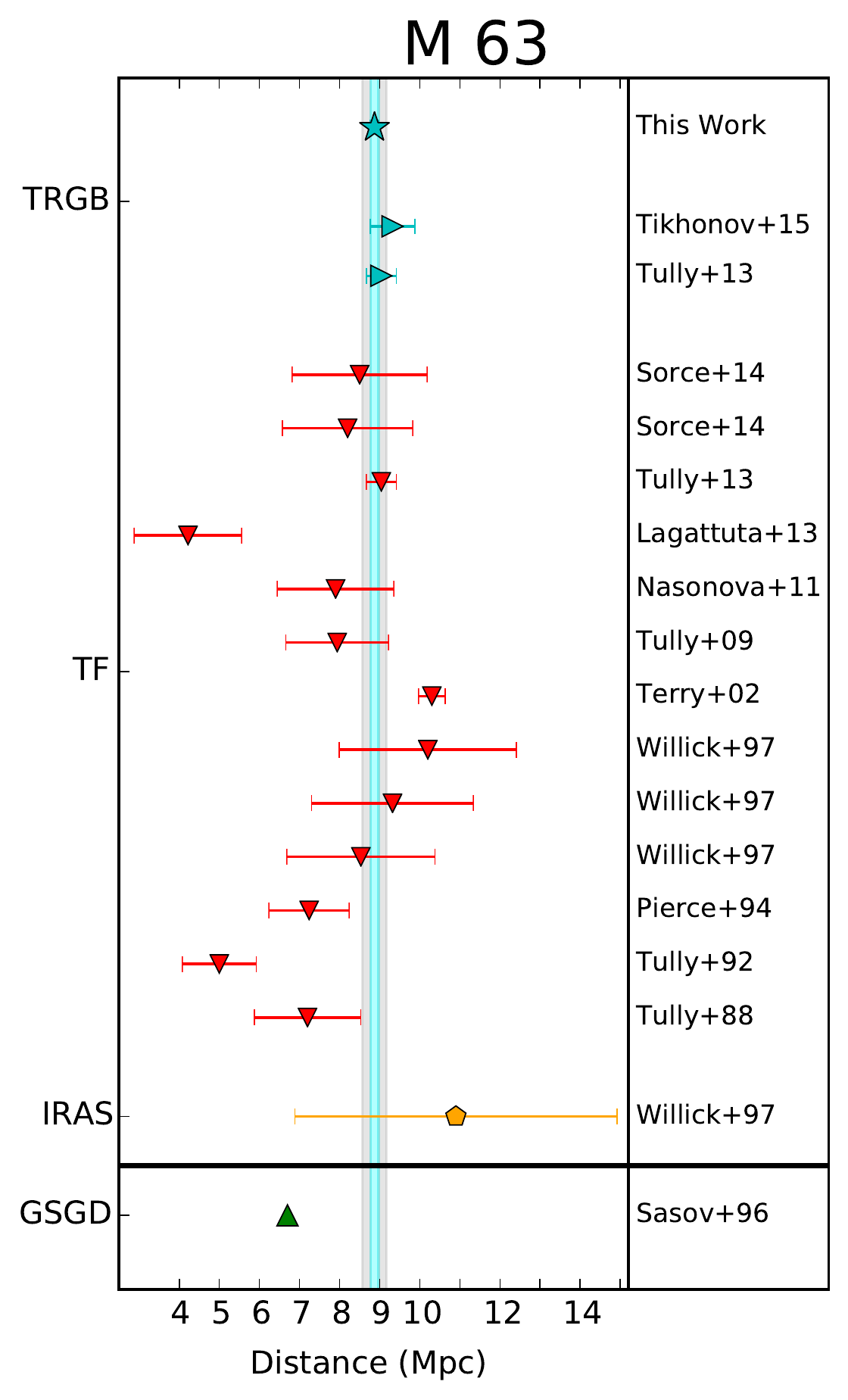}
\caption{Comparison of distance measurements to M63 from the literature with our value. To aid the comparison between our value and the previous values, we have added a shaded cyan vertical line centered on our TRGB measurements whose width encompasses the 1~$\sigma$ statistical uncertainty in our measurement and a grey line encompassing the 1~$\sigma$ statistical and systematic uncertainties added in quadrature. Distances from each distance method are given a unique color and plot symbol. See Table~\ref{tab:m63} for individual distance values, references, and sources for the data.}
\label{fig:m63}
\end{figure}

\begin{figure}
\includegraphics[width=\linewidth]{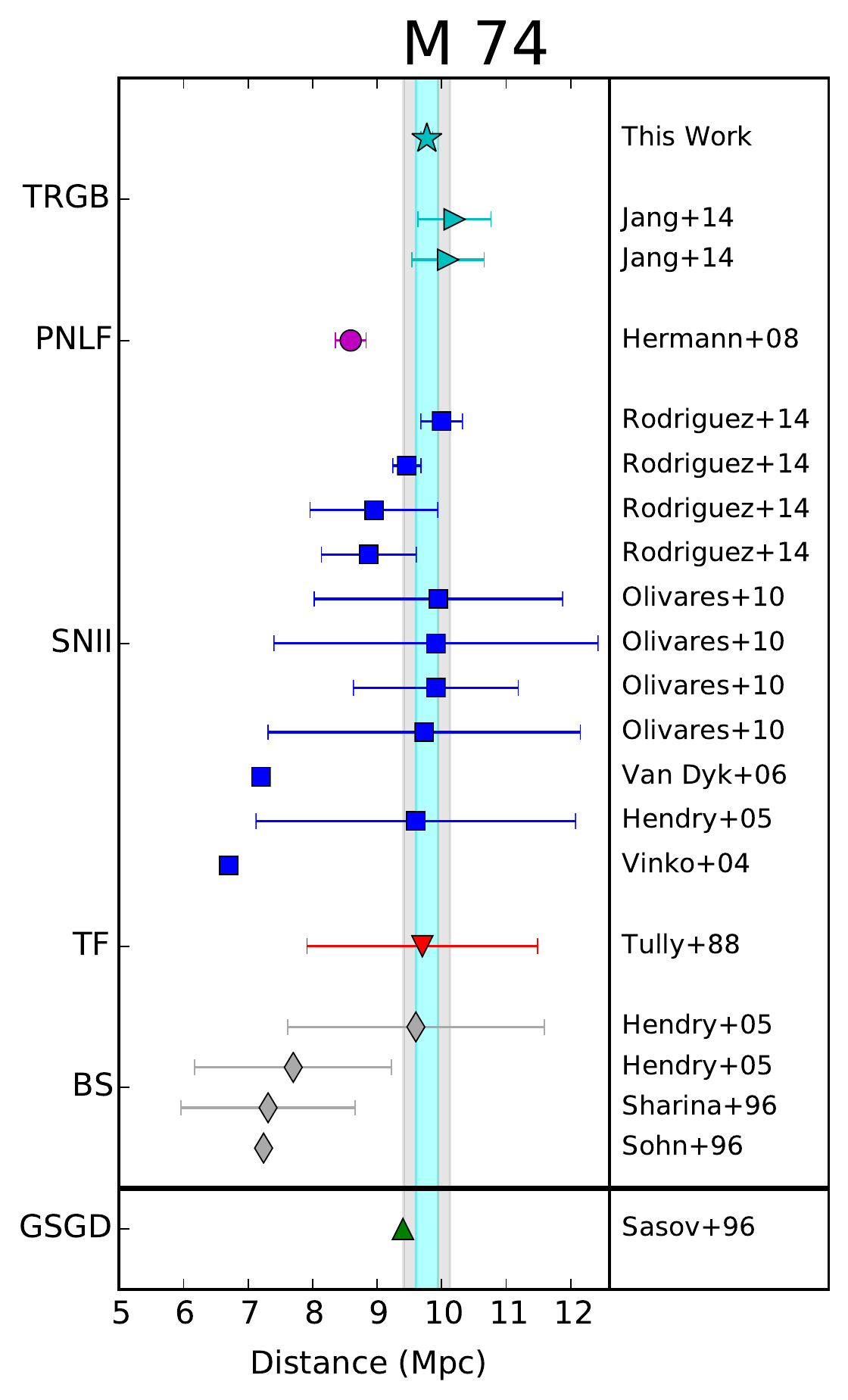}
\caption{Comparison of distance measurements to M74 from the literature with our value. The shaded lines correspond to the uncertainties in our TRGB measurement as described for Figure~\ref{fig:m63}. See Table~\ref{tab:m74} for individual distance values, references, and sources for the data.}
\label{fig:m74}
\end{figure}

Figures~\ref{fig:m63} $-$ \ref{fig:ngc5398} compare our TRGB distance measurements for each galaxy with previous distances measurements from the literature using various techniques. Tables~\ref{tab:m63} $-$ \ref{tab:ngc5398} provide the individual values, methods, references, and original data sources. Our distances are reported for each galaxy with the systematic uncertainties stated first, followed by the systematic uncertainties (see Section~\ref{sec:distances} for details). 

In the comparisons, we have included all publications found under the distance measurements in the NASA/IPAC Extragalactic Database (NED) for the sample. The distances from the literature are measured using a variety of techniques. In the Appendices of Paper~I and II, we provide descriptions of several methods including the surface brightness fluctuations (SBF), planetary nebulae luminosity function (PNLF), supernova Type II (SNII), Tully-Fisher (TF), CO ring (CO), gravitational stability of gaseous disks (GSGD), and the globular cluster luminosity function (GCLF). In the Appendix of this work, we provide descriptions of two additional methods, the brightest star (BS) and Infrared Astronomical Satellite Redshift Survey (IRAS) approaches. 

Note that some of the previous distance methodologies, such as the BS, GSGD, IRAS, SBF, or CO ring methods, have high uncertainties. In addition, multiple TF distances for an individual galaxy are available with results that can vary by a factor of 2 or more, or inappropriately determined for face-on spirals. These methods have been superseded by more accurate distance indicators such as the TRGB or Cepheid approachs. We include all distances to show the wide range of values in the literature, highlighting the need for careful discernment when adopting a distance value from NED.

%tab:m63
\begin{table*}
%McQuinn et al. 2016 Distance to Spirals: M63
\begin{center}
\caption{Distance Measurements to M~63}
\label{tab:m63}
\end{center}
\begin{center}
\vspace{-15pt}
\begin{tabular}{llll}
\hline 
dm (mag)			& D (Mpc)			& Reference			& Data	\\
\hline 
\hline
\\
\multicolumn{4}{c}{\textbf{Tip of the Red Giant Branch}}\\
29.74$\pm0.02\pm0.07$	& $8.87\pm0.10$$\pm0.29$ & This work 		& new  \\
\\
29.85	$\pm$0.13	& 9.32		& \citet{Tikhonov2015} 	& archival \\
29.78	$\pm$0.09	& 9.04		& \citet{Tully2013} 		& archival \\
\\
\multicolumn{4}{c}{\textbf{Tully-Fisher Relation}}\\
29.65	$\pm$0.43	& 8.50		& \citet{Sorce2014} 		& archival \\
29.58	$\pm$0.43	& 8.20		& \citet{Sorce2014} 		& archival \\
29.78	$\pm$0.09	& 9.04		& \citet{Tully2013} 		& archival \\
28.13	$\pm$0.69	& 4.22		& \citet{Lagattuta2013} 	& archival \\
29.49	$\pm$0.40	& 7.90		& \citet{Nasonova2011} 	& archival \\
29.50	$\pm$0.35	& 7.94		& \citet{Tully2009} 		& archival \\
30.07	$\pm$0.07	& 10.30		& \citet{Terry2002} 		& archival \\
29.66	$\pm$0.47	& 8.53		& \citet{Willick1997}  		& archival \\
29.85	$\pm$0.47	& 9.32		& \citet{Willick1997} 		& archival \\
30.03	$\pm$0.47	& 10.20		& \citet{Willick1997} 		& archival \\
29.30	$\pm$0.30	& 7.24		& \citet{Pierce1994} 		& new \& archival  \\
28.49	$\pm$0.40	& 5.00		& \citet{Tully1992} 		& archival \\
29.27	$\pm$0.40	& 7.20		& \citet{Tully1988} 		& archival \\
\\
\multicolumn{4}{c}{\textbf{IRAS redshift survey reconstruction method}}\\
30.20   $\pm$0.80       	& 10.90 		& \citet{Willick1997}   	& archival \\
\\
\hline
\multicolumn{4}{c}{\textbf{Gravitational Stability of Gas Disk}}\\
29.85				& 6.70		& \citet{Zasov1996}		& archival \\
\hline\hline
\end{tabular}
\end{center}
\tablecomments{Distance measurements from the literature from various techniques. The Reference column lists the source of the reported measurement. The Data column lists whether the observations were original to the study or were from data archives. Figure~\ref{fig:m63} shows the distribution of the measurements. Details on the TF relation and GSDS method can be found in the Appendix of Paper~I; a brief description of the IRAS redshift method is presented in the Appendix of this work.}

\end{table*}

In each of the Figures~\ref{fig:m63} $-$ \ref{fig:ngc5398}, we overlay a vertical shaded cyan line centered on our distance measurement with a width encompassing the 1$\sigma$ statistical uncertainty in distance, and a shaded grey line encompassing the 1$\sigma$ statistical and systematic uncertainty added in quadrature. The various previous distance measurements are grouped by technique and, within each technique, are listed from the most recent to the oldest. Some of the studies measured multiple distances based on the same data set. We list the separate, individual measurements to show the range found by the different studies, highlighting the challenges in measuring precise distances.

\subsection{The Distance to M63}
\begin{center}
{\it (NGC~5055; The Sunflower Galaxy)}
\end{center}
We measure the TRGB distance to M63 to be $8.87\pm0.10\pm0.29$ Mpc. Our value lies within the previously reported distance range spanning 5.0 $-$ 10.9 Mpc. Figure~\ref{fig:m63} compares our TRGB distance for M63 with the 17 distances from 12 different papers reported in NED based on the TRGB, TF, IRAS, and GSGD methods; individual values and references are listed in Table~\ref{tab:m63}.  

Our TRGB distance agrees with the recently reported TRGB distances from \citet{Tully2013} found in the Extragalactic Distance Database as part of the Cosmic-Flows-2 program and from \citet{Tikhonov2015}. The \citet{Tully2013} value is also based on a ML technique and calibrated using \citet{Rizzi2007a}. The \citet{Tikhonov2015} is based on a Sobel filter approach but the calibration was not specified. Our measurement has smaller uncertainties due to a number of factors including (i) the use of deeper photometry with smaller photometric uncertainties in the magnitude range of the TRGB,  (ii) the application of spatial cuts to the photometry to minimize contamination from non-RGB stars, and (iii) in the comparison with \citet{Tikhonov2015}, we use a ML technique with calibrations specific to the $HST$ filters. Thus, our measurement is an improvement over previous results. 

Thirteen of the distances to M63 are based on the TF relation varying from 5.0 $-$ 10.2 Mpc; nine of these overlap within the uncertainties of our value. Note that the TF distances include measurements based on adaptation of the original method \citep{Tully1988}. The majority are based on the traditional method using luminosities and \HI\ line widths from studies of the kinematics of the local universe. Additional attempts were made to improve the method by using infrared luminosities to trace the stellar mass \citep{Lagattuta2013, Sorce2014} and look-alike (or `sosies') galaxies that can be calibrated against one another \citep{Terry2002, Paturel1984}.

Distance measurements have also been obtained using IRAS-based redshifts and the GSGD method. The IRAS redshift survey estimates a distance with larger uncertainties, in part because distances based on recessional velocities can be skewed by peculiar local motions in the Local Volume. Finally, the GSGD method was an older attempt to find a stable geometric feature in spiral galaxies that could be used as a distance indicator. While no uncertainties were provided for this measurement, the GSGD approach has higher intrinsic uncertainties. We include it here for completeness. 

\subsection{The Distance to M74}
\begin{center}
{\it (NGC~628; archetype grand-design spiral)}
\end{center}

We measure the distance to M74 to be $9.77\pm0.17\pm0.32$ Mpc. Our value can be compared with twenty distances ranging from $6.7 - 10.2$ Mpc based on TRGB, PNLF, SN~II, TF, BS, and GSGD methods, shown in Figure~\ref{fig:m74} and listed in Table~\ref{tab:m74}.

Our TRGB distance is in agreement with two TRGB distances recently reported by \citet{Jang2014} using $HST$ archival imaging of M74. This study combined several short exposures of an overlapping region to reach the required photometric depth for a TRGB identification. Photometry was performed on the co-added images, rather than combining the stellar fluxes measured from the individual exposures. While both approaches will reach the same photometric depth, performing photometry on the combined mosaic image degrades the stable $HST$ point spread function and, thus, reduces the photometric accuracy. \citet{Jang2014} converted the $HST$ photometry to the Johnson V, I system and used a Sobel filter on stars selected in a $\sim0.5$ mag wide color interval to identify the discontinuity in the I-band equivalent LF. Our measurement is an improvement over this previous study as we use higher accuracy photometry in the native $HST$ filter set and a ML technique which takes into account the completeness of the data. 

M74 hosts three Type~II SNe which have been extensively studied and used as distance indicators to the galaxy. Most of the studies use the same photometric data obtained on the SNe events. Multiple distances are reported in individual studies based on applying different expanding photometric models to the data and correlating the magnitude of the SN with the expansion velocity of the photospheres. Our TRGB distance overlaps with nine of the eleven reported SN II distances. 

%tab:m74
\begin{table*}
%McQuinn et al. 2016 Distance to Spirals: M74
\begin{center}
\caption{Distance Measurements to M~74}
\label{tab:m74}
\end{center}
\begin{center}
\vspace{-15pt}
\begin{tabular}{llll}
\hline 
dm (mag)		& D (Mpc)	& Reference	& Data	\\
\hline 
\hline
\\
\multicolumn{4}{c}{\textbf{Tip of the Red Giant Branch}}\\
29.95$\pm0.03\pm0.07$	& 9.77$\pm0.17\pm0.32$ 	& This work 	& new  \\
\\
30.04$\pm0.03\pm0.12$	& 10.20	& \citet{Jang2014} 		& archival \\ 
30.02$\pm0.03\pm0.12$	& 10.10	& \citet{Jang2014} 		& archival \\ 
\\
\multicolumn{4}{c}{\textbf{Planetary Nebulae Luminosity Function}} \\
29.67   $\pm$0.45        	& 9.6        & \citet{Hermann2008} 	& new  \\ 
\\
\multicolumn{4}{c}{\textbf{Optical Type II Supernova}}\\
30.01	$\pm$0.07	& 10.00	& \citet{Rodriguez2014} 	& SN 2013ej \\
29.88	$\pm$0.05	& 9.46	& \citet{Rodriguez2014}	&  SN 2013ej \\
29.76	$\pm$0.24	& 8.95	& \citet{Rodriguez2014}	&  SN 2003gd \\
29.74	$\pm$0.18	& 8.87	& \citet{Rodriguez2014} 	&  SN 2003gd \\
29.99	$\pm$0.42	& 9.95	& \citet{Olivares2010} 	&  SN 2003gd \\
29.98	$\pm$0.55	& 9.91	& \citet{Olivares2010} 	&  SN 2003gd \\
29.98	$\pm$0.28	& 9.91	& \citet{Olivares2010} 	&  SN 2003gd \\
29.94	$\pm$0.54	& 9.73	& \citet{Olivares2010} 	&  SN 2003gd \\
29.29				& 7.20	& \citet{VanDyk2006} 	&  SN 2003gd \\
29.91	$\pm$0.56	& 9.60	& \citet{Hendry2005}		&  SN 2003gd \\
29.13				& 6.70	& \citet{Vinko2004}		&  SN 2002ap \\
\\
\multicolumn{4}{c}{\textbf{Tully-Fisher Relation}}\\
29.93	$\pm$0.40	& 9.700	& \citet{Tully1988} 		& archival \\
\\
\multicolumn{4}{c}{\textbf{Brightest Stars}} \\
29.91	$\pm$0.45	& 9.60	& \citet{Hendry2005} 	& new  and archival \\
29.43	$\pm$0.43	& 7.70	& \citet{Hendry2005} 	& new  and archival \\
29.32	$\pm$0.40	& 7.31	& \citet{Sharina1996} 	& new  \\
29.30				& 7.24	& \citet{Sohn1996}		& new  \\
\\
\hline
\multicolumn{4}{c}{\textbf{Gravitational Stability of Gas Disk}}\\
29.87				& 9.400	& \citet{Zasov1996}		& archival \\
\hline\hline
\end{tabular}
\end{center}
\vspace{-10pt}
\tablecomments{Distance measurements from the literature from various techniques. The Reference column lists the source of the reported measurement. The Data column lists whether the observations were original to the study or were from data archives, which sometimes included a re-calibration of existing work in the literature. In the case of the SN II distances, the individual SN event(s) used in each study is listed. Figure~\ref{fig:m74} shows the distribution of the measurements.  Details on the SBF, PNLF, SNII, TF, and GSDS methods can be found in the Appendix of Paper~I; a brief description of the BS method is presented in the Appendix of this work.}

\end{table*}

The BS method was employed by a number of studies who used the luminosity of blue supergiants \citep{Sharina1996} or blue and red supergiants \citep{Sohn1996} to determine the distance. In \citet{Sharina1996}, distances determined from blue supergiants in the dwarf irregular companions of M74 provided a consistency check on the M74 distance measurement. A more recent BS distance from \citet{Hendry2005} uses higher-resolution $HST$ imaging in an attempt to reduce uncertainties due to misidentifying stellar complexes, \HII\ regions, or foreground stars as single blue supergiants. The \citet{Hendry2005} study reported two distances; the farther 9.60 Mpc distance was favored by their final analysis. However, the method of using either blue or red supergiants as precision distance indicators was called into question in \citet{Rozanski1994}. Based on a comprehensive comparison of observational data, these authors estimate a minimum uncertainty of 0.55 mag in distance modulus for all calibration techniques due to a dependency on the parent galaxy's luminosity.

Additional individual measurements to M74 exist based on the PNLF, TF, and GSGD methods. The TF relation and GSGD method results overlap with our value, but the PNLF value does not.

\begin{figure}
\includegraphics[width=\linewidth]{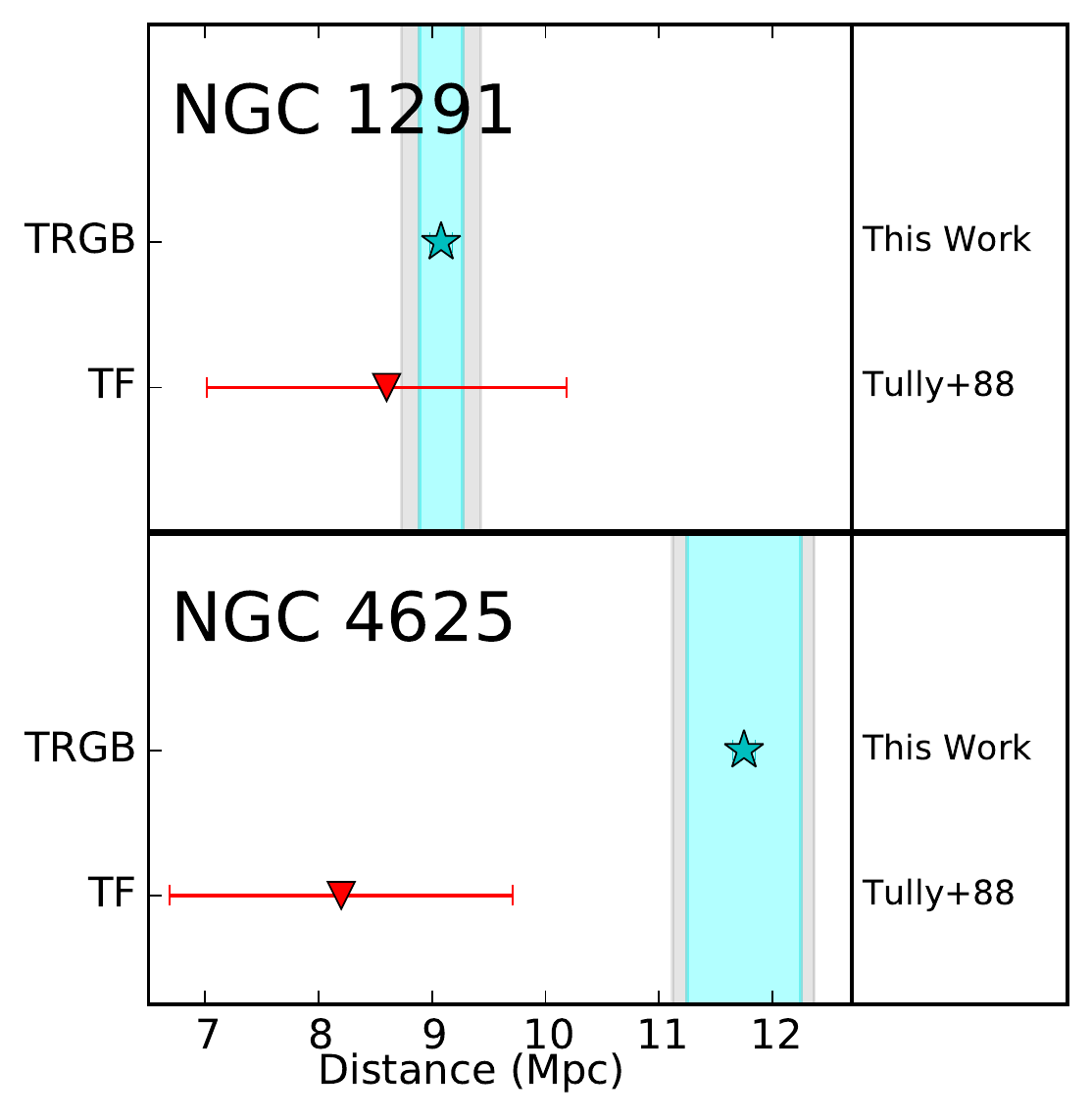}
\caption{Comparison of distance measurements to NGC~1291 and NGC~4625 from the literature with our value. The shaded lines correspond to the uncertainties in our TRGB measurement as described for Figure~\ref{fig:m63}. See Tables~\ref{tab:ngc1291} \& \ref{tab:ngc4625} for individual distance values, references, and sources for the data.}
\label{fig:ngc1291_ngc4625}
\end{figure}

\subsection{The Distance to NGC~1291 and NGC~4625}
We measure the TRGB distances to NGC~1291 to be $9.08\pm0.19\pm0.29$ Mpc and to NGC~4625 to be $11.75\pm0.50\pm0.38$ Mpc. Only one distance measurement was listed in NED for each galaxy based on the TF relation. These values are  shown in Figure~\ref{fig:ngc1291_ngc4625} and listed in Tables~\ref{tab:ngc1291} \& \ref{tab:ngc4625}. For NGC~1291, the TF distance estimate overlaps with our TRGB distance. For NGC~4625, the TF distance is smaller with no overlap in $1\sigma$ uncertainties with our TRGB distances. However, the TF distance likely has higher than stated uncertainties due to two factors. The galaxy is close to face on, which can add uncertainty to the inclination correction used in determining the TF distance. In addition, NGC~4625 is interacting with the nearby galaxy NGC~4618, adding uncertainty in using emission line widths to trace the mass of the galaxy in the TF relation. 

\begin{figure}
\includegraphics[width=\linewidth]{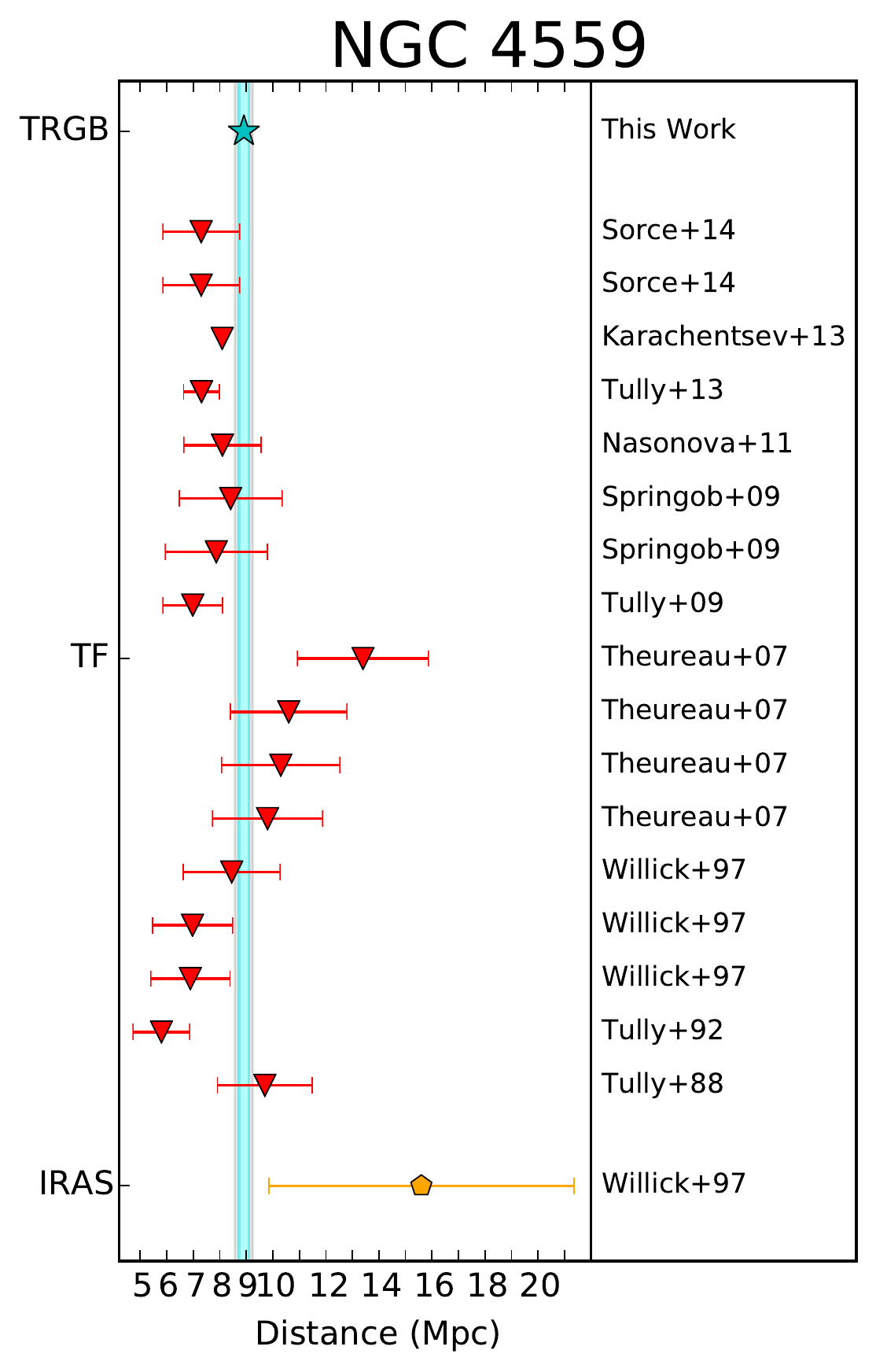}
\caption{Comparison of distance measurements to NGC~4559 from the literature with our value. The shaded lines correspond to the uncertainties in our TRGB measurement as described for Figure~\ref{fig:m63}. See Table~\ref{tab:ngc4559} for individual distance values, references, and sources for the data.}
\label{fig:ngc4559}
\end{figure}

\begin{figure}
\includegraphics[width=\linewidth]{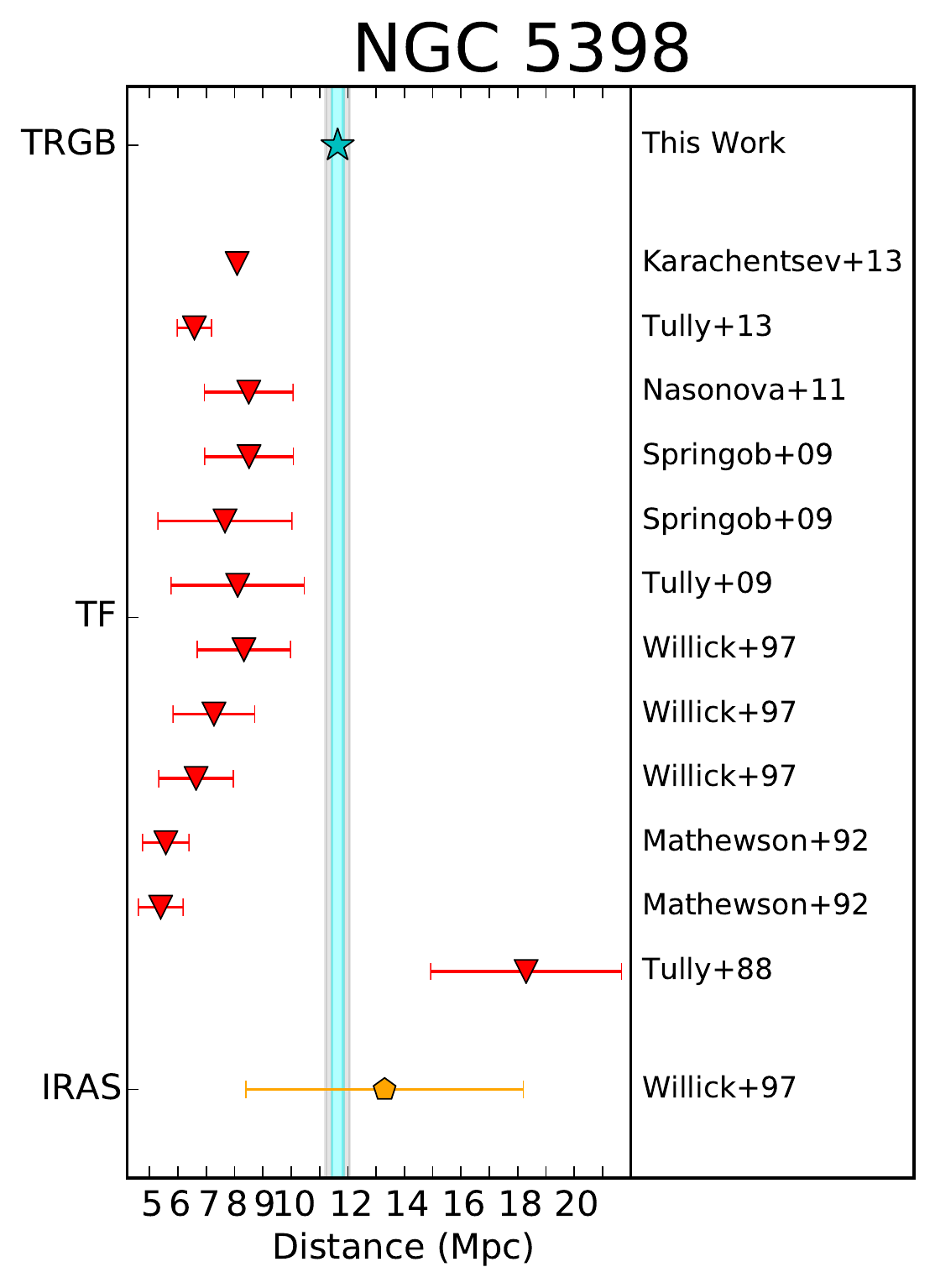}
\caption{Comparison of distance measurements to NGC~5398 from the literature with our value. The shaded lines correspond to the uncertainties in our TRGB measurement as described for Figure~\ref{fig:m63}. See Table~\ref{tab:ngc5398} for individual distance values, references, and sources for the data.}
\label{fig:ngc5398}
\end{figure}

\subsection{The Distance to NGC~4559}
We measure the TRGB distance to NGC~4559 to be $8.91\pm0.19\pm0.29$ Mpc. Our value lies within the previously reported distance range spanning 5.8 $-$ 13.4 Mpc. Figure~\ref{fig:ngc4559} compares our TRGB distance measurement to eighteen previous measurements from ten different studies; the individual distances are listed in Table~\ref{tab:ngc4559}. All but one measurement are based on the TF relation. Ten of the TF distances agree within the uncertainties with our measurement; more recent TF distances have favored a smaller distance. An additional distance was reported from the IRAS redshift survey, with large uncertainties likely due to the peculiar velocity flows in the nearby universe. 

\subsection{The Distance to NGC~5398}
We measure the TRGB distance to NGC~5398 to be $11.64\pm0.20\pm0.38$ Mpc. Our value lies within the previously reported distance range spanning 5.39 $-$ 18.30 Mpc. Figure~\ref{fig:ngc5398} compares our TRGB distance measurement to thirteen previous measurements from eight different studies; the individual distances are listed in Table~\ref{tab:ngc5398}. Almost all measurements are based on the TF relation, which do not agree with our measurement. An additional distance with large uncertainties was reported from the IRAS redshift survey. 
\vspace{15pt}

%tab:ngc1291
\begin{table}
%McQuinn et al. 2016 Distance to Spirals: NGC 1291
\begin{center}
\caption{Distance Measurements to NGC~1291}
\label{tab:ngc1291}
\end{center}
\begin{center}
\vspace{-15pt}
\begin{tabular}{llll}
\hline 
dm (mag)				& D (Mpc)			& Reference			& Data	\\
\hline 
\hline
\\
\multicolumn{4}{c}{\textbf{Tip of the Red Giant Branch}}\\
29.79$\pm0.05\pm0.07$		& $9.08\pm0.19\pm0.29$ 	& This work 		& new  \\
\\
\multicolumn{4}{c}{\textbf{Tully-Fisher Relation}}\\
29.68	$\pm$0.40	& 8.63			& \citet{Tully1988} 		& archival \\
\\
\hline\hline
\end{tabular}
\end{center}
\vspace{-10pt}
\tablecomments{Our TRGB distance compared with a distance measurement from the literature for NGC~1291. The top panel in Figure~\ref{fig:ngc1291_ngc4625} compares our measurement with the previous TF distance measurement. Details on the TF relation can be found in the Appendix of Paper~I.}

\end{table}

%tab:ngc4625
\begin{table}
%McQuinn et al. 2016 Distance to Spirals: NGC 4625
\begin{center}
\caption{Distance Measurements to NGC~4625}
\label{tab:ngc4625}
\end{center}
\begin{center}
\vspace{-15pt}
\begin{tabular}{llll}
\hline 
dm (mag)				& D (Mpc)			& Reference			& Data	\\
\hline 
\hline
\\
\multicolumn{4}{c}{\textbf{Tip of the Red Giant Branch}}\\
30.35$\pm0.09\pm0.07$	& $11.75\pm0.50\pm0.38$ & This work & new  \\
\\
\multicolumn{4}{c}{\textbf{Tully-Fisher Relation}}\\
29.57	$\pm$0.40	& 8.20$\pm1.51$	& \citet{Tully1988} 		& archival \\
\\
\hline\hline
\end{tabular}
\end{center}
\vspace{-10pt}
\tablecomments{Our TRGB distance compared with a distance measurement from the literature for NGC~4625. The bottom panel in Figure~\ref{fig:ngc1291_ngc4625} compares our measurement with the previous TF distance measurement. Details on the TF relation can be found in the Appendix of Paper~I.}

\end{table}

%tab:ngc4559
\begin{table*}
%McQuinn et al. 2016 Distance to Spiral Galaxies: NGC 4559 
\begin{center}
\caption{Distance Measurements to NGC~4559}
\label{tab:ngc4559}
\end{center}
\begin{center}
\vspace{-15pt}
\begin{tabular}{llll}
\hline 
dm (mag)				& D (Mpc)	& Reference				& Data	\\
\hline 
\hline
\\
\multicolumn{4}{c}{\textbf{Tip of the Red Giant Branch}}\\
29.75$\pm0.05\pm0.07$	& $8.91\pm0.19\pm0.29$ & This work 	& new \\
\\
\multicolumn{4}{c}{\textbf{Tully-Fisher Relation}}\\
29.32	$\pm$0.43	& 7.30	& \citet{Sorce2014} 			& archival \\
29.32	$\pm$0.43	& 7.30	& \citet{Sorce2014} 			& archival \\
29.54				& 8.09	& \citet{Karachentsev2013} 	& archival \\
29.32	$\pm$0.20	& 7.31	& \citet{Tully2013} 			& archival \\
29.54	$\pm$0.39	& 8.10	& \citet{Nasonova2011}  		& archival \\
29.63	$\pm$0.50	& 8.41	& \citet{Springbob2009} 		& archival \\
29.48	$\pm$0.53	& 7.87	& \citet{Springbob2009} 		& archival \\
29.22	$\pm$0.35	& 6.98	& \citet{Tully2009} 			& archival \\
30.64	$\pm$0.40	& 13.40	& \citet{Theureau2007} 		& new  \\ 
30.12	$\pm$0.45	& 10.60	& \citet{Theureau2007} 		& new  \\ 
30.06	$\pm$0.47	& 10.30	& \citet{Theureau2007} 		& new  \\ 
29.96	$\pm$0.46	& 9.80	& \citet{Theureau2007} 		& new  \\ 
29.64	$\pm$0.47	& 8.45	& \citet{Willick1997} 			& archival \\ 
29.22	$\pm$0.47	& 6.97	& \citet{Willick1997} 			& archival \\ 
29.19	$\pm$0.47	& 6.89	& \citet{Willick1997} 			& archival \\ 
28.82	$\pm$0.40	& 5.80	& \citet{Tully1992} 			& archival \\
29.93	$\pm$0.40	& 9.70	& \citet{Tully1988} 			& archival \\ 
\\
\multicolumn{4}{c}{\textbf{IRAS redshift survey reconstruction method}}\\
30.97	$\pm$0.80	& 15.60	& \citet{Willick1997} 			& archival \\
\\
\hline\hline
\end{tabular}
\end{center}
\vspace{-10pt}
\tablecomments{Distance measurements from the literature from various techniques. The Reference column lists the source of the reported measurement. The Data column lists whether the observations were original to the study or were from data archives. Figure~\ref{fig:ngc4559} shows the distribution of the measurements. Details on the TF relation can be found in the Appendix of Paper~I; a brief description of the IRAS redshift method is presented in the Appendix of this work.}

\end{table*}

%tab:ngc5398
\begin{table*}
%McQuinn et al. 2016 Distance to Spiral Galaxies: NGC 5398 
\begin{center}
\caption{Distance Measurements to NGC~5398}
\label{tab:ngc5398}
\end{center}
\begin{center}
\vspace{-15pt}
\begin{tabular}{llll}
\hline 
dm (mag)				& D (Mpc)	& Reference				& Data	\\
\hline 
\hline
\\
\multicolumn{4}{c}{\textbf{Tip of the Red Giant Branch}}\\
30.33$\pm0.04\pm0.07$	& $11.64\pm0.20\pm0.38$ & This work 	& new \\
\\
\multicolumn{4}{c}{\textbf{Tully-Fisher Relation}}\\
29.54				& 8.09	& \citet{Karachentsev2013} 	& archival \\
29.09	$\pm$0.20	& 6.58	& \citet{Tully2013} 			& archival \\
29.65	$\pm$0.40	& 8.50	& \citet{Nasonova2011}  		& archival \\
29.42	$\pm$0.67	& 7.66	& \citet{Springbob2009} 		& archival \\
29.55	$\pm$0.63	& 8.11	& \citet{Springbob2009} 		& archival \\
29.65	$\pm$0.40	& 8.51	& \citet{Tully2009} 			& archival \\
29.60	$\pm$0.43	& 8.33	& \citet{Willick1997} 			& archival \\
29.31	$\pm$0.43	& 7.27	& \citet{Willick1997} 			& archival \\
29.11	$\pm$0.43	& 6.64	& \citet{Willick1997} 			& archival \\
28.74	$\pm$0.32	& 5.57	& \citet{Mathewson1992} 		& new  \\ 
28.66	$\pm$0.32	& 5.39	& \citet{Mathewson1992}		& new  \\
31.31	$\pm$0.40	& 18.30	& \citet{Tully1988} 			& archival \\
\\
\multicolumn{4}{c}{\textbf{IRAS redshift survey reconstruction method}}\\
30.61	$\pm$0.80	& 13.30	& \citet{Willick1997} 			& archival \\
\\
\hline\hline
\end{tabular}
\end{center}
\tablecomments{Distance measurements from the literature from various techniques. The Reference column lists the source of the reported measurement. The Data column lists whether the observations were original to the study or were from data archives. Figure~\ref{fig:ngc5398} shows the distribution of the measurements. Details on the TF relation can be found in the Appendix of Paper~I; a brief description of the IRAS redshift method is presented in the Appendix of this work.}

\end{table*}

\section{Discussion}
Significant effort has been dedicated to accurately measuring distances to galaxies in the local universe using a variety of methods. Each method has its own set of challenges and uncertainties, and the TRGB method is no exception. Despite the theoretical `knife-edge' break in the $I-$band LF at the luminosity of the TRGB, the discontinuity can be broadened in the complex, composite stellar populations of spiral galaxies. Therefore, the most accurate measurement becomes dependent on a number of factors. As discussed above, selecting a field of view in the outer region of the system is critical to reduce crowding, contamination from non-RGB stellar populations, and potentially the metallicity spread of the RGB stars which can bias the measurement and/or increase uncertainties. The field selection at large galactocentric radii must be balanced by choosing a region that has a significant numbers of stars such that the upper RGB is not impacted by completeness issues, from not only the depth of photometry but also from random stochastic changes in the stellar populations. Applying extinction corrections and a color-based correction for the spread in metallicity prior to fitting for the TRGB increases the sharpness in the discontinuity, thereby improving the measurement. Using a ML technique to fit the LF can significantly decrease the uncertainties compared to a Sobel filter, given the broad peaks in the filter response, but the Sobel filter continues to provide a consistency check on the ML technique. Applying color-cuts to the photometry to select RGB is essential, but applying too narrow a color-range can bias the results. Systematic uncertainties for the TRGB method still vary. We have adopted the $1\sigma$ estimate of 0.07 mag from \citet{Rizzi2007a} based on the calibration of horizontal branch stars from \citet{Carretta2000} and the mean calibration of the Large Magellanic Cloud distances, but note that \citet{Bellazzini2001} estimate $1\sigma$ systematic uncertainties to be 0.12 mag using data on the globular cluster $\omega$ Centauri. Parallax measurements to both the Large Magellanic Cloud and globular clusters from the Gaia telescope will hopefully improve calibrations in the future.

Prior work comparing the PNLF and SBF methods have found that the PNLF distances are generally $\sim$15\% smaller than the SBF distances due to systematic errors in the calibration galaxies and reddening uncertainties \citep{Ciardullo2012}. These authors argue the `true' distance scale is somewhere in between the two methods. In the two galaxies in our study with PNLF, SBF, and TRGB distances, we find that PNLF distances to be smaller than the SBF distances in one galaxy (M104), with the TRGB distance in between. The remaining system (M51) shows an overlap between the PNLF and SBF distances, with the TRGB value slightly higher. 

For galaxies with multiple TF distances, we find a wide range in results not only between the TF and TRGB distances, but also between the TF distance measurements themselves. The TF relation is predicated on the premise that spiral galaxies have consistent mass to light ratios, therefore the rotational speed and luminosity are linearly correlated. However, luminosity also depends on the recent star formation history of a system, particularly at optical wavelengths. More recent TF distances use infrared luminosities intended to minimize the impact of recent changes in star formation history. In the galaxies studied here, this has sometimes resulted in distances that are closer to our TRGB measurement \citep[e.g., M63, NGC~4559;][]{Sorce2014}. The notable exception is M104, but the TF method is not a reliable distance method given its peculiar morphology and gas properties. Regardless, given the wide range in TF distances available for a given galaxy, TF distances are most useful for obtaining approximate distance estimates. Care should be taken when adopting a TF distance for precision calculations on nearby galaxies. 

Two galaxies in our sample have SNII distances that have been calculated numerous times using different calibrations and expanding photospheric models. Because of the assumptions needed to convert the luminosity and geometry of a core-collapse Type II SN into a standard-candle, the results have a wide dispersion. \citet{Hamuy2001} previously noted observational errors, dilution factors, velocity interpolations, and dust extinction contribute significantly to the uncertainties and report typical distance errors of order 20\% with a dispersion in distance modulus of $\sim0.2-0.4$ mag. For M51, the mean SNII distance from 19 results is 7.7 Mpc with a standard deviation of 1.0 Mpc (corresponding to $\sim0.3$ mag in distance modulus), compared with our TRGB distance of $8.58\pm0.10\pm0.$28 Mpc. Similarly for M74, the mean SNII distance from 11 results is 9.1 Mpc with a standard deviation of 1.2 Mpc, compared with our TRGB distance of $9.77\pm0.17\pm0.32$ Mpc. 

One galaxy in our sample, M104, has GCLF distances measured in two studies \citep{Bridges1992, Spitler2006}, as discussed in Paper~II. The reported distances span nearly 10 Mpc, ranging from $6.2-15.8$ Mpc. In a review of observational and theoretical studies on the GCLF method, \citet{Rejkuba2012} report that additional corrections are needed to account for galaxy type, environment, and dynamical history of the host galaxy. Without such corrections, the GCLF calibrations suffer from important systematic uncertainties. Given the complicated morphology of M104, GCLF based distances likely have higher, unquantified uncertainties than reported. 

\section{Conclusions}\label{sec:conclusions}
We have measured the TRGB distances to a sample of well-studied spiral galaxies in the nearby galaxies. In this paper, we report the distances to M63, M74, NGC~1291, NGC~4559, NGC~4625, and NGC~5398. We previously reported the distances to M51 \citep{McQuinn2016a} and M104 \citep{McQuinn2016b}. The full sample is part of the SINGS, KINGFISH, THINGS, NGS, HERACLES, and PINGS programs and four overlap with the CHAOS programs; many of the results of these programs depend on accurate distances. 

The distances are based on measuring the luminosity of the TRGB in $HST$ optical imaging of resolved stellar populations. We apply a ML technique to identify the TRGB, which parameterizes the luminosity function and takes into account photometric uncertainties and incompleteness. We also apply a Sobel filter as a consistency check on our results. We use the calibration from \citet{Rizzi2007a} to convert the TRGB luminosity to a distance modulus, which includes a color-based correction for metallicity and is specific to the $HST$ filter set.

We compare the distances to each galaxy in our sample to previously reported distances in the literature from a variety of techniques. Depending on the technique, there can be a wide range in published distances, particularly from the TF relation. In three galaxies, there have been other recently reported TRGB distances. While our measurements have smaller uncertainties driven mainly by deeper photometry and the use of the ML technique, the different TRGB distances are all in excellent agreement. Despite different measurement techniques (i.e., a Sobel filter vs. ML technique) and different data sets (different fields of view and photometric depths), the robust and well-calibrated TRGB method returns consistent distances. 

\acknowledgments
Support for this work was provided by NASA through grant GO-13804 from the Space Telescope Institute, which is operated by Aura, Inc., under NASA contract NAS5-26555. This research made use of NASA's Astrophysical Data System and the NASA/IPAC Extragalactic Database (NED) which is operated by the Jet Propulsion Laboratory, California Institute of Technology, under contract with the National Aeronautics and Space Administration. 

{\it Facilities:} \facility{Hubble Space Telescope}

\appendix
\section{Descriptions of Distance Indicators}\label{sec:appendix}
In Paper~I, we gave brief descriptions of the SBF, PNLF, SNII, TF, GSDS, and CO Ring methods used to measure the distance to M51. In Paper~II, we added a description of GCLF to the Appendix. Here, we add description of two additional metrics used to measure the distance in the remainder of our sample, namely the Brightest Stars (BS), and IRAS redshift survey reconstruction method (IRAS).  

\subsection{Brightest Stars (BS)}
Originally used by \citet{Hubble1936} for building an extragalactic distance scale, distances based on the brightest stars method assume that the brightest blue and red supergiants have a nearly constant visual luminosity. The accuracy of the method is driven by uncertainties in photometric scales, membership of the source to the galaxy, the potential for misidentifying compact, blue stellar complexes as single blue supergiants, and the dependence of the luminosity of the brightest stars with the luminosity of the galaxy \citep[][and references therein]{Piotto1992, Sharina1996, Karachentsev1994}. While some of these uncertainties can be mitigated with additional observational constraints (e.g., membership uncertainties can by mitigated by combining photometric and spectroscopic observations), \citet{Rozanski1994} report a minimum 1$\sigma$ uncertainty of 0.55 mag in the calibration of the BS method due a dependency on the parent galaxy's luminosity. 

\subsection{Infrared Astronomical Satellite Redshift Survey (IRAS)}
The Infrared Astronomical Satellite (IRAS) surveyed 87.6\% of the sky at 60$\micron$ down to a flux limit of 1.2 Jy. From the galaxies identified in the IRAS data, new and archival spectroscopic observations were used to determine the redshifts and map the position and velocity of the sample on the sky \citep{Strauss1992, Fisher1995}. The IRAS redshift survey then mapped the large-scale number density of galaxies. The redshifts can be converted to distances based on Hubble's Law. For systems with redshift velocities $< 2500$ km s$^{-1}$, the results show the well-known anisotropic distribution of galaxies caused by local over-densities and peculiar motions of galaxies in  local density fields. Thus, for any individual nearby galaxy, particularly for galaxies within the Local Volume, redshift-based distances can be inaccurate due to local galaxy flows and peculiar velocities impacting the conversion from redshift velocity to distance. 

\renewcommand\bibname{{References}}
\bibliography{../../bibliography.bib}

\end{document}